\documentclass[11pt]{article}
\usepackage{amsfonts,amscd,amsmath, amssymb,graphicx,color}
\newcommand{\oh}{\frac{1}{2}}

\def\ep{\text{e}}
\def\ap{\alpha'}

\def\4{\tfrac{1}{4}}

\title{Generating Functional for Gauge Invariant Actions: Examples of Nonrelativistic Gauge Theories}
\author{Oleg Andreev\thanks{Also at Landau Institute for Theoretical Physics, Moscow.}
\\ \\
{\it Arnold Sommerfeld Center for Theoretical Physics, LMU-M\"unchen,} \\
{\it Theresienstrasse 37, 80333 M\"unchen, Germany}}
\date{}

\begin{document}

\maketitle
\begin{abstract}
We propose a generating functional for nonrelativistic gauge invariant actions. In particular, we consider actions 
without the usual magnetic term. Like in the Born-Infeld theory, there is an upper bound to the electric field strength in 
these gauge theories.

\end{abstract}

\vspace{-10cm}
\begin{flushright}
LMU-ASC 42/09
\end{flushright}
\vspace{9cm}


\section{Introduction}
\renewcommand{\theequation}{1.\arabic{equation}}
\setcounter{equation}{0}
The Wilson loops \cite{wilson}, with ${\cal C}$ a closed curve, $A$ the gauge field, and the trace taken in the $N$-dimensional fundamental 
representation of $SU(N)$

\begin{equation}\label{wilson-loop}
W(A, {\cal C})=\frac{1}{N}\text{tr} P \exp\Bigl[ i\oint_{\cal C} dx^\mu A_\mu\Bigr]
\,,
\end{equation}
have been studied from many points of view in gauge theories \cite{reviews}. 

The first thought about loops is the Wilson criterion for confinement. It is expected to show up in an area law for the expectation value of 
a loop 

\begin{equation}\label{wilson-loop2}
{\cal W}({\cal C})=\int [dA] \,\ep^{-S_{\text{\tiny YM}}}\,\, W(A, {\cal C})
\,,
\end{equation}
with $S_{\text{\tiny YM}}$ the Yang-Mills action. The area law means that ${\cal W}({\cal C})$ decays exponentially with an area enclosed 
by the loop. There have also been many discussions of this issue in the context of supersymmetric gauge theories with additional 
fermionic and scalar fields \cite{reviews}.

The second thought about loops is that instead of quantizing the gauge field $A_\mu$, one can quantize coordinates $x^\mu$. In terms of 
Euclidean path integral, that means

\begin{equation}\label{Zg}
Z(A, {\cal C})=\int [d x]\,\ep^{-S(x)}\,\,W(A, {\cal C})
\,,
\end{equation}
where $S(x)$ is an action for the quantized coordinates, fields $x^\mu$ with $\mu=0,\dots,p$. For a $U(1)$ gauge group, such an 
integral appeared in the Feynman's first-quantized formulation of scalar QED  many years ago \cite{feynman}. Later, it also appeared 
in string theory, where the non-Abelian Wilson factor \eqref{wilson-loop} was inserted into the Polyakov path integral to describe an effective 
action for massless string modes \cite{fts}. The main point here is that the renormalized path integral \eqref{Zg} is to be identified with the 
effective action\footnote{It may need some refinement in the presence of scalar fields. For a discussion see \cite{tachyon} and references therein.}

\begin{equation}\label{Z=S}
S(A)=Z(A)
\,.
\end{equation}
Subsequent work \cite{bi,ts-rev} has made it clear that this approach should be taken seriously, particularly in the context of $D$-brane actions \cite{polchinski}.

In fact, one can consider the formula \eqref{Zg} as a generating functional for gauge invariant actions without any reference to string theory.
In other words, given the action $S(x)$, one can use it to compute (generate) the corresponding gauge theory action via \eqref{Zg}.
Note that in general such actions include infinitely many terms. The two classical examples are that of  Schwinger \cite{schwinger} and 
non-linear electrodynamics of Born-Infeld type obtained from \eqref{Zg} in \cite{bi}.\footnote{Some extrapolation formulas can be found in appendix of \cite{ts-rev}.} 

In this paper we have two main aims. First, we want to generalize the method to nonrelativistic gauge theories. In recent years, such theories 
have been of considerable interest in the context of strongly correlated electron systems as an effective field theory description of the 
long-wavelength physics at quantum critical points with anisotropic scaling. A peculiar property of those is that in the Lagrangian the usual 
$\mathbf{H}^2$-term \cite{fradkin}, or alternatively the $\mathbf{E}^2$-term \cite{freedman}, has vanishing coefficient. For example, in 
$3+1$ dimensions the action is of the form \cite{fradkin}

\begin{equation}\label{huse}
S(\mathbf{E},\mathbf{H})=\int d^3 x dt \left[\mathbf{E}^2-\rho_4(\nabla\times\mathbf{H})^2\right]
\,,
\end{equation}
where $\mathbf{E}^2=\mathbf{E}\cdot\mathbf{E}$ and $\rho_4$ is a coupling constant parameterizing a line of fixed points. 

Our second aim is to find some (off-critical) non-linear deformations of the effective actions proposed in \cite{fradkin}. We keep in mind 
the Born-Infeld electrodynamics \cite{bornI} that differs from that of Maxwell by higher order terms in the field strength. Moreover, 
as in the case of the Born-Infeld action where it describes the low energy dynamics of $D$-branes, such deformations might be useful 
for studying nonrelativistic branes.

Before getting to the specific examples that we will consider, let us set the basic framework. For the contour ${\cal C}$, we take a unit circle 
parameterized by an angular variable $\varphi$. Then, following \cite{fts}, we split the field ${x}^\mu(\varphi)$ into a zero Fourier mode $x^\mu_0$ and 
remaining non-zero modes such that $x^\mu(\varphi)=x^\mu_0+\xi^\mu(\varphi)$ with $\int_0^{2\pi}\varphi\,\xi^\mu=0$.\footnote{In what follows we omit 
the index $0$ from the zero mode when it is clear from the context.} A good pragmatic 
reason for doing so is that the $x$'s are interpreted as classical coordinates, while the $\xi$'s are quantum fluctuations which have to be 
integrated out. If we let $S(x)$ be quadratic in fluctuations, then, for a generic kinetic term, we can rewrite \eqref{Zg} as

\begin{equation}\label{Zg2}
S(A)=\int dx\int [d\xi]\,\exp\left[-\tfrac{1}{2}\xi G^{-1}\xi\right]\,  W(A(x+\xi))
\,,
\end{equation}
where $\xi G^{-1}\xi=\iint_0^{2\pi} d\varphi_1d\varphi_2 \,\xi^\mu (\varphi_1) G^{-1}_{\mu\nu}(\varphi_1,\varphi_2)
\xi^\nu (\varphi_2)$. We normalize the functional integral measure $[d\xi]$ as $\int [d\xi]\,\exp\bigl[ -\tfrac{1}{2}\xi G^{-1}\xi\bigr]=1$. 

In conclusion, we make the following remarks about formula \eqref{Zg2}:  

(1) For $S(x)=\tfrac{1}{2}\int_0^{2\pi} d\varphi\,\dot x\dot x$ with $\dot x=\partial_\varphi x$
that describes the free propagation of a particle, $G^{-1}$ is a local operator. Its Green function is given by 
$G(\varphi_{12})=\frac{1}{\pi}\sum_{n=1}^\infty\frac{1}{n^2}\cos n\varphi_{12}$, where $\varphi_{12}=\varphi_1 -\varphi_2$. 

(2) For the "string'' action $S(x)=\tfrac{1}{2}\int_D d^2\sigma\,\partial^a x^\mu\partial_a x_\mu$, with $D$ a unit disc, $G^{-1}$ is a non-local 
operator. Note that non-locality appears after Gaussian integration over $x(\sigma)$ in all internal points of the disc that provides an effective 
one-dimensional path integral \cite{bi}. In this case, the corresponding Green function is given by $G(\varphi_{12})=\tfrac{1}{\pi}\sum_{n=1}^\infty\frac{1}{n}\cos n\varphi_{12}$. 

(3) Finally, there is one more issue to be mentioned here. The above construction seems somewhat formal that may be when the path integral diverges. Following \cite{nahm}, we will define one-dimensional path integral by using the Riemann $\zeta$-function. Thus, we express all sums in terms of 
$\zeta(s)=\sum_{n=1}^\infty \tfrac{1}{n^s}$ and $\zeta(s,\tfrac{1}{2})=\sum_{r=1/2}^\infty\tfrac{1}{r^s}$. 

\section{Gauge Invariant Action without Usual Magnetic Term}
\renewcommand{\theequation}{2.\arabic{equation}}
\setcounter{equation}{0}
In this section we will describe a concrete example of nonrelativistic gauge theory without the usual magnetic $F_{ij}^2$-term in the Lagrangian. 
Since the Wilson factor \eqref{wilson-loop} is Lorentz invariant, we choose a non-invariant kinetic term

\begin{equation}\label{non-r}
S(x)=\frac{1}{2}\int_0^{2\pi}d\varphi
\Bigl[\frac{1}{(2\pi\alpha')^2}\dot x^0\dot x^0+\frac{1}{2\pi\alpha'\kappa}(x^i-x_0^i)(x^i-x_0^i)\Bigr]	
\,,
\end{equation}
or, equivalently,

\begin{equation}\label{kinetic}
\xi G^{-1}\xi=\frac{1}{(2\pi\alpha')^2}\xi^0 G^{-1}_{00}\xi^0+\frac{1}{2\pi\alpha'\kappa}\xi^i G^{-1}_{ij}\xi^j
\,.
\end{equation}
Here
\begin{equation}\label{propagators}
G^{00}(\varphi_{12})={\text G}_{12}=\frac{1}{\pi}\sum_{n=1}^\infty\frac{1}{n^2}\cos n\varphi_{12}
\,,\quad
G^{ij}(\varphi_{12})=\delta^{ij}\mathbf{G}_{12}
=\frac{1}{\pi}\delta^{ij}\sum_{n=1}^\infty\cos n\varphi_{12}
\,,
\end{equation}
where $\int_0^{2\pi}d\varphi_2\, G_{12}G^{-1}_{23}=\delta^+(\varphi_{13})$. The function $\delta^+(\varphi)$ plays a role of 
the $\delta$-function on 
the $\xi$'s. Therefore, it is given by $\delta^+(\varphi)=\frac{1}{\pi}\sum_{n=1}^\infty\cos n\varphi$. The indices $i$,$j$ run from $1$ to $p$. 
Like in string theory, we use $2\pi\alpha'$ for dimensional purposes. The normalization is designed to describe the gauge theory action with anisotropic 
scaling characterized by the dynamical critical exponent $z=2$. In this case dimensions of all objects are measured in the units of spatial momenta 
such that $\text{dim}\,x^0=-2$ and $\text{dim}\,x^i=-1$. $\kappa$ is a relative factor of dimension zero. With such a choice, we define the zero 
mode measure as $dx=\tau_p dt d^p x$, where $t=x^0$, $\tau_p^{-1}=g (2\pi\alpha')^{(p+2)/2}$ and $g$ is a dimensionless parameter.\footnote{In string theory $g$ is proportional to the string coupling.} This provides $\text{dim}\,dx=0$.

Remark (1) above makes it clear that fluctuations in time direction are described by the usual particle term. The integral \eqref{Zg2} for this case 
was intensively discussed in the literature.\footnote{For a review, see, e.g., \cite{particle} and references therein.} On the other hand, fluctuations in spacial directions are described by a sum of terms such that every term represents a white Gaussian noise.\footnote{Another possible way is to interpret them as a quadratic open string tachyon profile. However, a crucial difference with \cite{witten} is that the standard 
$\partial X^i\bar\partial X_i$-term is missing in our case.} To our knowledge, there have been no studies of \eqref{Zg2} with the noise term in the literature.

\subsection{Leading Terms}

To actually compute leading terms in the $\alpha'$-expansion of \eqref{Zg2}, one would have to expand $A(x+\xi)$ in powers of $\xi$ 
and then compute a few Feynman diagrams shown in Figure 1.\footnote{The diagrams look like those of \cite{N} because in both cases the vertices are 
the same.}

\vspace{1.25cm}
\begin{figure}[ht]
\centering
\includegraphics[width=12cm]{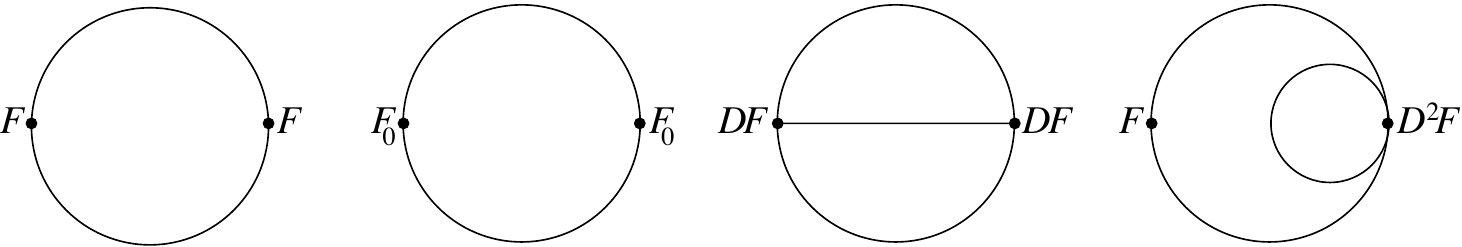}
\caption{\small{Feynman diagrams that contribute the leading terms in the $\alpha'$-expansion. $F_0$, $F$, $D$ stand for $F_{0i}$, $F_{ij}$, $D_i$, respectively.}}
\end{figure}

\vspace{.75cm}
With $\zeta(-2)=0$, the result, up to third order in $\alpha'$, is given by 

\begin{equation}\label{A1}
\begin{gathered}
S(A)=\tau_p\int dt d^px\,\text{tr}
\Bigl[1
+(2\pi\ap)^3\Bigl(b_1 F_{0i}^2+
b_2 (D_iF_{ij})^2+b_3 F_{ij}F_{jk}F_{ki}\Bigr)
+O(\ap^4)
\Bigr]
\,,\\
b_1=-\kappa\zeta(0)
\,,\quad
b_2=-\frac{1}{2\pi}\kappa^3\zeta^2(-1)
\,,\quad
b_3=-\frac{3}{2}i\kappa^3\zeta^2(-1)
\,.
\end{gathered}
\end{equation}
Here $D_\mu=\partial_\mu-iA_\mu$ is the covariant derivative and $F_{\mu\nu}=i[D_\mu,D_\nu]$ is the field strength. With $\zeta(-1)=-1/12$ and $\zeta(0)=-1/2$, the above equation becomes  

\begin{equation}\label{Z2}
S(A)=\tau_p\int dt d^px\,\text{tr}
\Bigl[1+4\kappa (\pi\alpha')^3 \Bigl( F_{0i}^2
-\frac{1}{144\pi}\kappa^2(D_iF_{ij})^2
-\frac{1}{48}i\kappa^2 F_{ij}F_{jk}F_{ki}\Bigr)
+O(\ap^4)
\Bigr]
\,.
\end{equation}
In the Abelian case the $F^3$ term vanishes and therefore \eqref{Z2} reduces to 

\begin{equation}\label{Z3}
S(A)=\tau_p\int dt d^px\,
\Bigl[1+4\kappa (\pi\alpha')^3 \Bigl( F_{0i}^2
-\frac{1}{144\pi}\kappa^2(\partial_iF_{ij})^2\Bigr)
+O(\ap^4)
\Bigr]
\,.
\end{equation}

An important remark about \eqref{Z2} and \eqref{Z3} is that the $\alpha'$-expansion doesn't coincide 
with a derivative expansion.

Motivated by the zero slope limit $(\alpha'\rightarrow 0)$ of string theory, we would like to consider it in the problem at hand. As known, 
the role of this limit is just to remove the higher order corrections in $\alpha'$ from the string effective action and, as a result, it becomes 
quadratic in $F$. 

So we take the limit $\alpha'\sim\varepsilon\rightarrow 0$, $g\sim\varepsilon^{2-(p/2)}$ with $\kappa$ held fixed. Ignoring the constant term we get 

\begin{equation}\label{huse2}
S(A)=\rho	
	\int dt d^px\,\text{tr}
\Bigl[F_{0i}^2
-\frac{1}{144\pi}\kappa^2(D_iF_{ij})^2
-\frac{1}{48}i\kappa^2 F_{ij}F_{jk}F_{ki}
\Bigr]
\,,
\end{equation}
where $\rho=\frac{\kappa}{2g}(2\pi\alpha')^{2-(p/2)}$. In contrast to string theory, the resulting action now includes the $DFDF$ and
$F^3$-terms. It becomes purely quadratic in the field strength only in the Abelian case when the cubic term identically vanishes.

Finally, it remains to be seen how the action \eqref{huse} is recovered. For this, we take $p=3$ and use a description in terms of 
ordinary (Abelian) electric and magnetic fields: $\mathbf{E}_i=F_{0i}$ and $\mathbf{H}_k=\frac{1}{2}\epsilon_{kij}F_{ij}$. Then a little 
algebra shows that \eqref{Z3} does reduce to \eqref{huse} with $\rho_4=-\frac{\kappa^2}{144\pi}$.\footnote{In the process, one has to rescale the 
electric field $\mathbf{E}\rightarrow i\mathbf{E}$ by the Wick rotation.}

\subsection{Slowly Varying Fields}

We now turn to the question of what deformations of the action \eqref{huse} are natural in our approach. Since \eqref{huse} describes the 
long wavelength physics, it is natural to consider the case of slowly varying, but not necessarily small, fields. For this we 
will ignore higher derivative terms in the $\xi$ expansion of $A(x+\xi)$. 

For slowly varying Abelian fields, \eqref{Zg2} reduces to 

\begin{equation}\label{BI}
S(A)=\int dx\int [d\xi]\,\exp\Bigl[ -\oh\xi G^{-1}\xi+
i\int^{2\pi}_0 d\varphi\, 
\dot\xi^j
\Bigl(F_{0j}\xi^0+\oh F_{ij}\xi^i
+\frac{1}{3}\partial_k F_{ij}\xi^i\xi^k
+\frac{1}{8}\partial_l\partial_k F_{ij}\xi^i\xi^k\xi^l
\Bigr)
\Bigr]
\,.
\end{equation}

It is convenient to integrate over $\xi^0$ first. Then, we get

\begin{equation}\label{BI2}
\begin{split}
S(A)=\int dx\int [d\xi^i]\,
\exp\Bigl[& -\frac{1}{2}\xi^i\Bigl(\frac{1}{2\pi\alpha'\kappa}\delta_{ij}\mathbf{G}^{-1}
+(2\pi\alpha')^2F_{0i}F_{0j}\ddot{\text G}\Bigr)\xi^j\\
& +
i\int^{2\pi}_0 d\varphi\, 
\xi^i\dot\xi^j
\Bigl(
\oh F_{ij}
+\frac{1}{3}\partial_k F_{ij}\xi^k
+\frac{1}{8}\partial_l\partial_k F_{ij}\xi^k\xi^l
\Bigr)
\Bigr]
\,,
\end{split}
\end{equation}
where $\ddot{\text G}=\partial_1\partial_2{\text G}(\varphi_{12})$.

As in string theory \cite{ats}, it is possible to include the $F^2\xi\xi$ and $F\dot\xi\xi$-terms into the propagator. However, for what follows we will 
include only one of them. To this end, we use the identity 

\begin{equation}\label{propagators-id}
\partial_1\partial_2\text{G}(\varphi_{12})=\mathbf{G}^{-1}(\varphi_{12})
\,
\end{equation}
to rewrite \eqref{BI2} as 

\begin{equation}\label{BI22}
S(A)=\int dx\int [d\xi^i]\,
\exp\Bigl[-\frac{1}{4\pi\alpha'\kappa}\xi^i{\cal G}_{ij}\mathbf{G}^{-1}\xi^j
+
i\int^{2\pi}_0 d\varphi\, 
\xi^i\dot\xi^j
\Bigl(
\oh F_{ij}
+\frac{1}{3}\partial_k F_{ij}\xi^k
+\frac{1}{8}\partial_l\partial_k F_{ij}\xi^k\xi^l
\Bigr)
\Bigr]
\,,
\end{equation}
where a new metric is given by

\begin{equation}\label{BI3}
{\cal G}_{ij}=\delta_{ij}+(2\pi\alpha')^3\kappa F_{0i}F_{0j}
\,.
\end{equation}
Note that ${\cal G}$ depends only on the electric field. 

We are now ready to compute a few terms in $S(A)$ by treating all the magnetic terms under the integral in \eqref{BI22} as perturbations. 
With the metric ${\cal G}$ held fixed, this is equivalent to a perturbation theory in $\alpha'$. Up to third order, the result is 

\begin{equation}\label{BI4} 
S(A)=\tau_p\int dtd^px\,\sqrt{\text{det}(\delta_{ij}+(2\pi\alpha')^3\kappa F_{0i}F_{0j})}
\biggl[1
-\frac{1}{36\pi}(\pi\alpha'\kappa)^3 \partial_n F_{ij}\partial_m F_{kl}{\cal G}^{ni}{\cal G}^{mk}{\cal G}^{jl}
+O(\alpha'^4)\biggr]
\,.
\end{equation}
Since we ignore derivatives of $F_{0i}$, we replaced the covariant derivatives (with respect to the metric ${\cal G}_{ij}$) with ordinary ones. 

To compare \eqref{BI4} to \eqref{Z3}, it is sufficient to expand $\sqrt{\text{det}\,{\cal G}}$ and ${\cal G}^{-1}$ in $\alpha'$. Since we are only interested in the leading terms, we substitute $\sqrt{\text{det}\,{\cal G}}=1+4(\pi\alpha')^3\kappa F_{0i}^2$ and ${\cal G}^{ij}=\delta^{ij}$ into \eqref{BI4} that immediately leads to the desired result. 

Now, specializing to four dimension and ignoring the higher $\alpha'$-corrections, we have

\begin{equation}\label{BI5} 
S(A)=\tau_3\int dtd^3x\,\sqrt{\text{det}(\delta_{ij}+(2\pi\alpha')^3\kappa\mathbf{E}_i\mathbf{E}_j)}
\biggl[1
-\frac{1}{36\pi}(\pi\alpha'\kappa)^3 
\partial_n\mathbf{F}_{ij}
\partial_m\mathbf{F}_{kl}
{\cal G}^{ni}{\cal G}^{mk}{\cal G}^{jl}
\biggr]
\,,
\end{equation}
with ${\cal G}_{ij}=\delta_{ij}+(2\pi\alpha')^3\kappa\mathbf{E}_i\mathbf{E}_j$ and $\mathbf{F}_{ij}=\epsilon_{ijk}\mathbf{H}_k$. It can be interpreted as a non-linear deformation of the original action \eqref{huse} by a slowly varying electric field. 

To illustrate the deformation \eqref{BI5}, we extend the leading order expression in the limit $\alpha'\rightarrow 0$ (again with $\mathbf{E}\rightarrow i\mathbf{E}/\sqrt{\rho}$ and $\mathbf{H}\rightarrow\mathbf{H}/\sqrt{\rho}$) to next order, and find  

\begin{equation}\label{huse3}
\begin{split}
{\cal S}=\int d^3 x dt \biggl[ &\mathbf{E}^2-\rho_4(\nabla\times\mathbf{H})^2
+\rho_\alpha\Bigl( 
\frac{1}{2}(\mathbf{E}^2)^2+\rho_4(\nabla\times\mathbf{H})^2\mathbf{E}^2 \\
& -2\rho_4(\mathbf{E}\cdot\nabla\times\mathbf{H})^2
-4\rho_4(\nabla\times\mathbf{H})\cdot(\mathbf{E}\times(\mathbf{E}\cdot\nabla)\mathbf{H})
\Bigr)
\biggr]
\,,
\end{split}
\end{equation}
where $\rho_\alpha=g(2\pi\alpha')^{\frac{5}{2}}$.

Finally, let us consider higher order terms in $F_{ij}$. We have already found that there is no $F_{ij}^2$ in the action. 
To extend this calculation to order $F^{2n}$, we must compute a connected Feynman diagram shown in Fig.2.\footnote{Note that in the Abelian case 
there are no terms which are odd in $F$.}

\vspace{1cm}
\begin{figure}[ht]
\centering
\includegraphics[width=2.5cm]{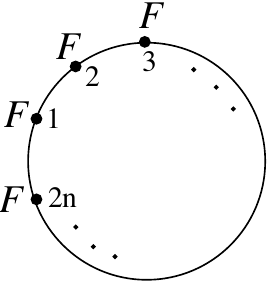}
\caption{\small{A Feynman diagram that contributes the $F^{2n}$-term to the action. $F$ stands for $F_{ij}$.}}
\end{figure}

\vspace{.5cm}
It is easy to see that this diagram is proportional to the following integral

\begin{equation}\label{M-diagram}
I_{2n}=\int_0^{2\pi}d\varphi_1\dots\int_0^{2\pi}d\varphi_{2n}\,\dot{\mathbf{G}}_{12}\dots\dot{\mathbf{G}}_{2n\,1}
\,,
\end{equation}
where $\dot{\mathbf{G}}_{km}=\partial_k\mathbf{G}_{nm}$. Integrating over the $\varphi$'s, we get 

\begin{equation}\label{integral}
I_{2n}=2(-)^n\zeta(-2n)=0
\,.
\end{equation}
We have used that $\zeta(-2n)=0$ for any positive integer $n$. So, there are no $F{\cal G}^{-1}F\dots F{\cal G}^{-1}$-terms in the action and, 
as a consequence, there are no $(\mathbf{H}^2)^n$-terms for $p=3$. This is the reason for not including the $F\dot\xi\xi$-term into the propagator in 
\eqref{BI22}.

\subsection{Some Generalizations}

We conclude this section with a computation that includes a more general form of the kinetic term. So we consider 

\begin{equation}\label{pr-ab}
{\text G}_{12}=\frac{1}{\pi}\sum_{n=1}^\infty\frac{1}{n^a}\cos n\varphi_{12}
\,,\quad
\mathbf{G}_{12}
=\frac{1}{\pi}\sum_{n=1}^\infty\frac{1}{n^b}\cos n\varphi_{12}
\,,
\end{equation}
where $a$ and $b$ are free parameters. Note that for $a=2$ and $b=0$, \eqref{pr-ab} reduces to \eqref{propagators}.

To compute the $F_{ij}^2$-term in the action, one has to evaluate the Feynman diagram as that of Fig. 2 with $n=1$ and the 
propagator defined by \eqref{pr-ab}. The integral \eqref{integral} becomes in this case

\begin{equation}\label{F2m}
I_2(b)=-2\zeta(2b-2)
\,.
\end{equation}

No essentially new computation is required to get the $F_{i0}^2$-term. Replacing $F_{ij}\xi^i\dot\xi^j$ by $F_{0j}\xi^0\dot\xi^j$ leads to 
the following integral

\begin{equation}\label{integral-e}
{\cal I}_2(a,b)=\int_0^{2\pi}d\varphi_1\int_0^{2\pi}d\varphi_2\,\dot{\mathbf{G}}_{12}\dot{\text G}_{21}
\,,
\end{equation}
which is simply

\begin{equation}\label{F2e}
{\cal I}_2(a,b)=-2\zeta(a+b-2)
\,.
\end{equation}

Now we are ready to see what vanishing the $F_{ij}^2$-term or alternatively the $F_{0i}^2$-term in the action means. From \eqref{F2m} it follows that 
there is no $F_{ij}^2$ if $2b-2$ coincides with one of the zeros of the Riemann zeta-function. For the trivial zeros, this restricts $b$ to be a non-positive integer. For the non-trivial zeros, $b$ is complex. It is given by $\frac{5}{4}+i\frac{t}{2}$.\footnote{According to the Riemann hypothesis, all the non-trivial zeros lie on the critical line $\frac{1}{2}+it$.} Similarly, the coefficient in front of the $F_{0i}^2$-term vanishes if $a+b-2$ is a zero of the zeta-function. 
This gives $a+b$ to be $2(1-k)$ with $k$ a positive integer for the trivial zeros and $\frac{5}{2}+it$ for the non-trivial zeros.

\section{Examples with Fermions}
\renewcommand{\theequation}{3.\arabic{equation}}
\setcounter{equation}{0}

The purpose of the present section is to add worldline fermions to the theory and do some explicit calculations illustrating our approach. 

With the fermions, the Wilson factor \eqref{wilson-loop} is typically extended to 

\begin{equation}\label{super-wilson}
\hat W(A, {\cal C})=\frac{1}{N}\text{tr} P \exp\Bigl[ i\int_0^{2\pi} d\varphi 
\Bigl(\dot x^\mu A_\mu-\frac{1}{2}F_{\mu\nu}\psi^\mu\psi^\nu\Bigr)\Bigr]
\,,
\end{equation}
such that it respects the supersymmetry transformations 
\begin{equation}\label{susy}
\delta x^\mu=\psi^\mu\epsilon
\,,\quad
\delta\psi^\mu=\dot x^\mu\epsilon
\,,
\end{equation}
even with a non-constant parameter $\epsilon$.

If we formulate the theory in the superfield notation, the Wilson factor \eqref{super-wilson} can be written in the 
following form \cite{ats1}

\begin{equation}\label{super-wilson1}
\hat W(A, {\cal C})=\frac{1}{N}\text{tr} 
\Bigl(\sum_{n=0}^\infty i^n\prod_{i=1}^n \int d\hat\varphi_i\, {\cal D}\hat x^\mu A_\mu (\hat x)
\prod_{j=1}^{n-1} \Theta(\hat\varphi_{j\,j+1})
\Bigr)
\,.
\end{equation}
Here 

\begin{equation}\label{superfield}
\hat x^\mu=x^\mu+\theta\psi^\mu
\,,\quad
{\cal D}=\theta\partial_\varphi -\partial_\theta
\,,\quad
d\hat\varphi=d\varphi d\theta
\,,\quad
\Theta(\hat\varphi_{ij})=\Theta(\varphi_{ij})+\theta_i\theta_j\delta(\varphi_{ij})
\,,
\end{equation}
where $\Theta$ is a step function.

Given the supersymmetry transformations, one can extend the action \eqref{non-r} to 

\begin{equation}\label{supernon-r}
\hat S(x,\psi)=\frac{1}{2}\int_0^{2\pi}d\varphi
\Bigl[\frac{1}{(2\pi\alpha')^2}\bigl(\dot x^0\dot x^0+\psi^0\dot\psi^0\bigr)+
\frac{1}{2\pi\alpha'\kappa}\bigl(x^i x^i-\psi^i\partial^{-1}\psi^i\bigr)
\Bigr]
\,
\end{equation}
by adding the fermions. Formally, it is invariant under \eqref{susy}. We consider for a moment the field $x^i$ rather than its projection on 
non-zero modes. The fact that $\psi\partial^{-1}\psi$ is a non-local operator is not in contradiction with our approach. The only restriction 
on $S(x,\psi)$ is that it must be quadratic in fluctuations.
   
Now we impose the antiperiodic boundary conditions on the fermions: $\psi^\mu(\varphi+2\pi)=-\psi^\mu(\varphi)$. This implies that there are no 
fermionic zero modes and, therefore, the $\psi$'s can be interpreted as additional quantum fluctuations. Then it is natural to combine bosonic 
and fermionic fluctuations into a single superfield 
$\hat\xi^\mu=\xi^\mu+\theta\psi^\mu$ such that $\hat x^\mu=x^\mu_0+\hat\xi^\mu$. 

With this choice of the boundary conditions we can write \eqref{supernon-r}, with $x^i_0$ subtracted from $x^i$, as  

\begin{equation}\label{superkin}
\xi G^{-1}\xi+\psi K^{-1}\psi=\frac{1}{(2\pi\alpha')^2}
\bigl(\xi^0 G^{-1}_{00}\xi^0+\psi^0 K^{-1}_{00}\psi^0\bigr)+ 
\frac{1}{2\pi\alpha'\kappa}
\bigl(\xi^i G^{-1}_{ij}\xi^j+\psi^i K^{-1}_{ij}\psi^j\bigr)
\,.
\end{equation}
Here
\begin{equation}\label{superpro1}
K^{00}(\varphi_{12})={\text K}_{12}=\frac{1}{\pi}\sum_{r=1/2}^\infty\frac{1}{r}\sin r\varphi_{12}
\,,\quad
K^{ij}(\varphi_{12})=\delta^{ij}\mathbf{K}_{12}
=\frac{1}{\pi}\delta^{ij}\sum_{r=1/2}^\infty r\sin r\varphi_{12}
\,.
\end{equation}
where $\int_0^{\pi}d\varphi_2\, K_{12}K^{-1}_{23}=\delta^-(\varphi_{13})$. The function $\delta^-(\varphi)$ plays a role of the $\delta$-function on 
the $\psi$'s. Explicitly, it is given by $\delta^-(\varphi_{12})=\frac{1}{\pi}\sum_{r=1/2}^\infty\cos r\varphi_{12}$.

In terms of $\hat\xi$, \eqref{superkin} reads

\begin{equation}\label{superkin1}
	\hat\xi \hat G^{-1}\hat\xi
	=\frac{1}{(2\pi\alpha')^2}\hat\xi^0\hat G^{-1}_{00}\hat\xi^0+
	\frac{1}{2\pi\alpha'\kappa}\hat\xi^i \hat G^{-1}_{ij}\hat\xi^j
\,.
\end{equation}
Here

\begin{equation}\label{superpro2}
\hat G^{00}(\varphi_{12})=\hat{\text{G}}_{12}=
{\text G}_{12}-\theta_1\theta_2{\text K}_{12}
\,,\quad
\hat{G}^{ij}(\varphi_{12})=\delta^{ij}\hat{\mathbf{G}}_{12}
=\delta^{ij}\bigl(\mathbf{G}_{12}-\theta_1\theta_2\mathbf{K}_{12}\bigr)
\,,
\end{equation}
where $\hat\xi\hat{G}^{-1}\hat\xi=
\iint d\hat\varphi_1d\hat\varphi_2 \,\hat\xi^\mu (\varphi_1)\hat{G}^{-1}_{\mu\nu}(\varphi_1,\varphi_2)
\hat\xi^\nu (\varphi_2)$ 
and $\int d\hat\varphi_2\, \hat{G}_{12}\hat{G}^{-1}_{23}=\theta_1\delta^-(\varphi_{13})-\theta_3\delta^+(\varphi_{13})$.\footnote{Note that 
$\hat{G}_{12}^{-1}=K^{-1}_{12}-\theta_1\theta_2 G_{12}^{-1}$.}

Once $\hat S$ is found, we will use it in the next subsection to study the generating functional 

\begin{equation}\label{super-Z}
S(A)=\int dx\int [d\hat\xi]\,\exp\left[-\tfrac{1}{2}\hat\xi \hat{G}^{-1}\hat\xi\right]\,  \hat{W}(A(x+\hat\xi))
\,.
\end{equation}
Here we normalize the functional integral measure $[d\hat\xi]$ as $\int [d\hat\xi]\,\exp\bigl[ -\tfrac{1}{2}\hat\xi \hat{G}^{-1}\hat\xi\bigr]=1$.

\subsection{Leading Terms}

We will now carry out a precisely analogous computation as that of subsection 2.1. The simplest way to do so is to use the superfield formalism. 
In practice, this means to replace $d\varphi$, $\Theta(\varphi)$, $\partial_\varphi$, $G$, etc. with $d\hat\varphi$, $\Theta(\hat\varphi)$, ${\cal D}$, $\hat G$, etc. in the corresponding expressions for the Feynman diagrams shown in Figure 1. Using $\zeta(-2)=\zeta(-2,\tfrac{1}{2})=0$, the diagrams 
can be evaluated to give 

\begin{equation}\label{S1}
\begin{gathered}
S(A)=\tau_p\int dt d^px\,\text{tr}
\Bigl[1
+(2\pi\ap)^3
\Bigl(s_1 F_{0i}^2+s_2 (D_iF_{ij})^2+s_3 F_{ij}F_{jk}F_{ki}\Bigr)
+O(\ap^4)
\Bigr]
\,,\\
s_1=-\kappa\bigl(\zeta(0)-\zeta(0,\tfrac{1}{2})\bigr)
\,,\quad
s_2=-\frac{1}{2\pi}\kappa^3\bigl(\zeta(-1)-\zeta(-1,\tfrac{1}{2})\bigr)^2
\,,\quad
s_3=-\frac{3}{2}i\kappa^3\bigl(\zeta(-1)-\zeta(-1,\tfrac{1}{2})\bigr)^2
\,.
\end{gathered}
\end{equation}
With $\zeta(0,\tfrac{1}{2})=0$ and $\zeta(-1,\tfrac{1}{2})=1/24$, \eqref{S1} takes the form

\begin{equation}\label{S2}
S(A)=\tau_p\int dt d^px\,\text{tr}
\Bigl[1+4\kappa (\pi\alpha')^3 \Bigl( F_{0i}^2
-\frac{1}{64\pi}\kappa^2(D_iF_{ij})^2
-\frac{3}{64}i\kappa^2 F_{ij}F_{jk}F_{ki}\Bigr)
+O(\ap^4)
\Bigr]
\,.
\end{equation}

A few noteworthy facts are the following. The effect of the fermions on the coefficients $b_i$ is a shift: $\zeta(s)\rightarrow\zeta(s)-\zeta(s,\tfrac{1}{2})$. The only exception is the coefficient in front of the quadratic term.
In fact, it doesn't get shifted because $\zeta(s,\tfrac{1}{2})=0$ for $s=0$. In the Abelian case the $F^3$ term vanishes. After taking 
the limit $\alpha'\rightarrow 0$ and setting $p=3$, the action reduces to \eqref{huse} with $\rho_4=-\frac{\kappa^2}{64\pi}$.


\subsection{Slowly Varying Fields}

Now let us explore some issues that arise in the case of slowly varying Abelian fields. We use again the superfield formalism.

For this case, \eqref{super-Z} reduces to

\begin{equation}\label{BIs}
S(A)=\int dx\int [d\hat\xi]\,
\exp\Bigl[ -\oh\hat\xi\hat{G}^{-1}\hat\xi+
i\int d\hat\varphi\, 
{\cal D}\hat\xi^j
\Bigl(F_{0j}\hat\xi^0+\oh F_{ij}\hat\xi^i
+\frac{1}{3}\partial_k F_{ij}\hat\xi^i\hat\xi^k
+\frac{1}{8}\partial_l\partial_k F_{ij}\hat\xi^i\hat\xi^k\hat\xi^l
\Bigr)
\Bigr]
\,.
\end{equation}

Integrating out $\hat\xi^0$, one gets simply

\begin{equation}\label{BIs2}
\begin{split}
S(A)=\int dx\int [d\xi^i]\,
\exp\Bigl[& -\frac{1}{2}\hat\xi^i\Bigl(\frac{1}{2\pi\alpha'\kappa}\delta_{ij}\hat{\mathbf{G}}^{-1}
-(2\pi\alpha')^2F_{0i}F_{0j}{\cal D}_1{\cal D}_2\hat{\text G}\Bigr)\hat\xi^j\\
& +
i\int d\hat\varphi\, 
\hat\xi^i{\cal D}\hat\xi^j
\Bigl(
\oh F_{ij}
+\frac{1}{3}\partial_k F_{ij}\hat\xi^k
+\frac{1}{8}\partial_l\partial_k F_{ij}\hat\xi^k\hat\xi^l
\Bigr)
\Bigr]
\,,
\end{split}
\end{equation}
where $\hat\xi{\cal D}_1{\cal D}_2\hat{\text{G}}\hat\xi=\int\int d\hat\varphi_1d\hat\varphi_2\,\hat\xi_1{\cal D}_1{\cal D}_2\hat{\text{G}}_{12}\hat\xi_2$.

It is easy to see now that 

\begin{equation}\label{id-s}
{\cal D}_1{\cal D}_2\hat{\text{G}}(\varphi_{12})=-\hat{\mathbf{G}}^{-1}(\varphi_{12})
\,
\end{equation}
is a superspace version of \eqref{propagators-id}. The minus sign is due to the fact that the ${\cal D}$'s
anticommute with each other. This identity allows us to combine the two first terms of \eqref{BIs2} into a single one. So, we have

\begin{equation}\label{BIs22}
S(A)=\int dx\int [d\xi^i]\,
\exp\Bigl[-\frac{1}{4\pi\alpha'\kappa}\hat\xi^i{\cal G}_{ij}\hat{\mathbf{G}}^{-1}\hat\xi^j
+
i\int d\hat\varphi\, 
\hat\xi^i{\cal D}\hat\xi^j
\Bigl(
\oh F_{ij}
+\frac{1}{3}\partial_k F_{ij}\hat\xi^k
+\frac{1}{8}\partial_l\partial_k F_{ij}\hat\xi^k\hat\xi^l
\Bigr)
\Bigr]
\,,
\end{equation}
with ${\cal G}_{ij}$ defined in \eqref{BI3}. 

Having derived \eqref{BIs22}, we are now in a position to compute a few terms in the action. Just as in section 2.2, keeping the 
metric ${\cal G}$ fixed, we find up to third order in $\alpha'$

\begin{equation}\label{BIs4} 
S(A)=\tau_p\int dtd^px\,\sqrt{\text{det}(\delta_{ij}+(2\pi\alpha')^3\kappa F_{0i}F_{0j})}
\biggl[1
-\frac{1}{16\pi}(\pi\alpha'\kappa)^3 \partial_n F_{ij}\partial_m F_{kl}{\cal G}^{ni}{\cal G}^{mk}{\cal G}^{jl}
+O(\alpha'^4)\biggr]
\,.
\end{equation}

To get from \eqref{BIs4} to \eqref{S2}, its Abelian version, we must expand ${\cal G}$ in $\alpha'$. In doing so, the $F_{0i}^2$ term comes from 
the determinant of ${\cal G}$, while the $(DF)^2$ term is already in \eqref{BIs4}. 

We conclude this section by computing the higher order terms in $F_{ij}$. For this, we need to evaluate the Feynman diagram of Figure 2 in the 
superfield formalism. This diagram is given, up to a constant multiple, by a multiple integral 

\begin{equation}\label{Ens1}
\hat I_{2n}=\int d\hat\varphi_1\dots\int d\hat\varphi_{2n}\,{\cal D}_1\hat{{\mathbf{G}}}_{12}\dots{\cal D}_{2n}\hat{{\mathbf{G}}}_{2n\,1}
\,.
\end{equation}
Performing integration, we obtain

\begin{equation}\label{Ens2}
I_{2n}=2(-)^{n+1}\bigl(\zeta(-2n)-\zeta(-2n,\tfrac{1}{2})\bigr)
=0
\,.
\end{equation}
We have used that $\zeta(-2n)=\zeta(-2n,\tfrac{1}{2})=0$ for any positive integer $n$. Thus, there are no $F{\cal G}^{-1}F\dots F{\cal G}^{-1}$ 
terms in the action and, as a consequence, there are no $(\mathbf{H}^2)^n$-terms for $p=3$.

\subsection{Some Further Generalizations}

To generalize the discussion of section 2.3 to the case with the fermions, the first step is to pick up fermionic kinetic terms such that

\begin{equation}\label{super-ab}
{\text K}_{12}=\frac{1}{\pi}\sum_{r=1/2}^\infty\frac{1}{r^{a-1}}\sin r\varphi_{12}
\,,\qquad
\mathbf{K}_{12}
=\frac{1}{\pi}\sum_{r=1/2}^\infty\frac{1}{r^{b-1}}\sin r\varphi_{12}
\,.
\end{equation}
So for $a=2$ and $b=0$, \eqref{super-ab} coincides with \eqref{superpro1}. Then, the corresponding superspace propagators are constructed by 
combining them with the $G$'s of section 2.3.

Now, to compute a coefficient in front of the $F_{ij}^2$-term in the action, one has to evaluate the Feynman diagram of Figure 2 with $n=1$. In the 
case of interest the integral \eqref{Ens1} becomes

\begin{equation}\label{F2sb}
\hat I_2(b)=2\bigl(\zeta(2b-2)-\zeta(2b-2,\tfrac{1}{2})\bigr)=4\bigl(1-2^{2b-3}\bigr)\zeta(2b-2)
\,.
\end{equation}
Here we have used the identity that $\zeta(s,\tfrac{1}{2})=(2^s-1)\zeta(s)$.

Similarly, one can find a coefficient in front of the $F_{0i}^2$-term. Replacing $F_{ij}\hat\xi^i{\cal D}\hat\xi^j$ by 
$F_{0j}\hat\xi^0{\cal D}\hat\xi^j$ leads to the following integral  

\begin{equation}\label{Es}
\hat{\cal I}_2(a,b)=\int d\hat\varphi_1\int d\hat\varphi_2\,{\cal D}_1\hat{{\mathbf{G}}}_{12}
{\cal D}_2\hat{{\text G}}_{21}
\,.
\end{equation}
After performing integration, one gets

\begin{equation}\label{Es2}
\hat{{\cal I}}_2(a,b)=2
\bigl(\zeta(a+b-2)-\zeta(a+b-2,\tfrac{1}{2})\bigr)=4\bigl(1-2^{a+b-3}\bigr)\zeta(a+b-2)
\,.
\end{equation}

Clearly, no new analysis is required to conclude what vanishing the $F_{ij}^2$ and $F_{0i}^2$-terms in the action means. 
Therefore our conclusions here are similar to those of section 2.3.  

\section{Concluding Comments}
\renewcommand{\theequation}{4.\arabic{equation}}
\setcounter{equation}{0}

We will here conclude with brief observations about our approach and the non-relativistic gauge invariant actions derived above. 

(1) For $F_{ij}=0$, the Abelian actions \eqref{BI4} and \eqref{BIs4} reduce to  

\begin{equation}\label{E}
S(A)=\tau_p\int dtd^px\,\sqrt{\text{det}(\delta_{ij}+F_{0i}F_{0j})}
\,.
\end{equation}
In order to keep the equations simple we have set $(2\pi\alpha')^3\kappa=1$. Clearly, the $\alpha'$-dependence can be easily restored on 
dimensional grounds. It is surprising that $\eqref{E}$ can be written in the Born-Infeld form 

\begin{equation}\label{E1}
S(A)=\tau_p\int dtd^px\,\sqrt{\text{det}(\delta_{\mu\nu}+F_{\mu\nu})}
\,.
\end{equation}
To show this, we derive a useful identity. We do that in a frame where $F_{\mu\nu}$ has a special form

\begin{equation}\label{E2}
F=
\begin{pmatrix}
	0&F_{01}&0&\dots &0\\
	-F_{01}&0&0&\dots &0\\
	0&\hdotsfor{3}&0\\
	\hdotsfor{5}\\
	0&\hdotsfor{3}&0
\end{pmatrix}
\,,
\end{equation}
with $F_{01}$ the only nontrivial component. Then, we find

\begin{equation}\label{E3}
\text{det}(\delta_{ij}+F_{0i}F_{0j})=\text{det}(\delta_{\mu\nu}+F_{\mu\nu})=1+F_{0i}^2
\,.
\end{equation}

The relation with the Born-Infeld theory implies that, after Wick rotating to Lorentzian signature, there is an upper bound to the electric field 
strength. If one restores dimensions, the critical field strength $E_c$ is given by 	

\begin{equation}\label{E4}
E_c=\frac{1}{\sqrt{(2\pi\alpha')^3\kappa}}
\,.
\end{equation}
Note that, as in string theory, there is no upper bound in the limit $\alpha'\rightarrow 0$. 

These similarities with the Born-Infeld action suggest a possibility of considering the actions we found above as candidates for effective actions of 
nonrelativistic branes.

(2) As noted above, in the presence of fermions $\zeta(s)$ combines with $\zeta(s,\tfrac{1}{2})$ to form a linear combination 
$\zeta(s)-\zeta(s,\tfrac{1}{2})$. It is very special in that it is a holomorphic function of the complex variable $s$. A simple pole of 
$\zeta(s)$ at $s=1$ cancels with that of $\zeta(s,\tfrac{1}{2})$. Indeed, one can easily check that  $\lim_{s\rightarrow 1}(\zeta(s)-\zeta(s,\tfrac{1}{2}))=-\ln 4$. This means that the Feynman diagrams we considered in section 3 remain finite for all 
values of $a$ and $b$. As a result, the path integral representation for the action \eqref{super-Z} is well-defined at least up to third order in 
$\alpha'$.

(3) Although the Gaussian noise terms describing fluctuations in spatial directions were introduced by hand so as to recover finally the gauge theory action \eqref{huse}, there is some indirect evidence for this. A few facts are particularly interesting: 

(i) This gauge theory action was proposed in \cite{fradkin} to describe the Rokhsar-Kivelson points of quantum dimer models. 

(ii) Recently, it was claimed in \cite{dij} that the quantization scheme used for the quantum dimer models is nothing else but a discrete analog of stochastic quantization.

(iii) As known, the notion of noise plays a pivotal role in stochastic quantization. 

If so, then it seems natural to expect the appearance of ''noise'' in our approach too. As we have shown above this is indeed the case.    
 
(4) It is worth mentioning that a non-Abelian gauge theory action without the usual magnetic term was also proposed in \cite{horava}.\footnote{Note that a non-Abelian action without the usual electric term was proposed in \cite{freedman}.} It differs from ours \eqref{huse2} and \eqref{S2} by the absence of the $F^3$-term. From our approach we don't see any reason for dropping it. It disappears in the open superstring effective 
action ($a=b=1$), where, on the other hand, the magnetic term is present because of Lorentz invariance. 

(5) So far, stringy motivated methods were developed only for relativistic field theories \cite{particle}. In this paper we have taken a first 
step to develop them for nonrelativistic gauge theories by deriving some gauge invariant actions. Clearly, this avenue of research deserves to be pursued. In addition to string theory, it might be quite important for nonrelativistic QCD, statistical and solid state physics.

\section{Acknowledgments}
We are grateful to A.A. Tseytlin for discussions and M. Haack for reading the manuscript. This work was supported in part by DFG within 
the Emmy-Noether-Program under Grant No.HA 3448/3-1, Excellence Cluster, and the Alexander von Humboldt Foundation under Grant No.PHYS0167. 
We also would like to thank D. Kharzeev and I. Zaliznyak for hospitality at the Brookhaven National Laboratory, where a portion of this 
work was completed.


\end{document}